\documentclass[final,authoryear]{elsart5p}
\usepackage{graphicx}
\usepackage{natbib}
\usepackage[dvipsnames,usenames]{color}
\let\oldsqrt\sqrt       
\def\sqrt{\mathpalette\DHLhksqrt}
\def\DHLhksqrt#1#2{%
\setbox0=\hbox{$#1\oldsqrt{#2\,}$}\dimen0=\ht0
\advance\dimen0-0.2\ht0
\setbox2=\hbox{\vrule height\ht0 depth -\dimen0}%
{\box0\lower0.4pt\box2}}

\begin{document}
\begin{frontmatter}
\title{Invasions in heterogeneous habitats in the presence of advection} 
\author[IAC]{Davide Vergni}
\author[RM]{Sandro Iannaccone} 
\author[FR2]{Stefano Berti} 
\author[ISC]{Massimo Cencini}

\address[IAC]{Istituto per le Applicazioni del Calcolo, CNR,  Via dei Taurini, 19 00185 Rome, Italy}

\address[RM]{Dipartimento di Fisica, University of Rome ``Sapienza'', Piazzale A. Moro 5, 00185 Rome, Italy}

\address[FR2]{Laboratoire de M\'et\'eorologie Dynamique, IPSL, ENS/CNRS, Paris, France} 

\address[ISC]{Istituto dei Sistemi Complessi, CNR, Via dei Taurini, 19 00185 Rome, Italy} 


\begin{abstract}
 We investigate invasions from a biological reservoir to an initially
 empty, heterogeneous habitat in the presence of advection. The
 habitat consists of a periodic alternation of favorable and
 unfavorable patches. In the latter the population dies at fixed
 rate. In the former it grows either with the logistic or with an
 Allee effect type dynamics, where the population has to overcome a threshold
 to grow. We study the conditions for successful invasions and the
 speed of the invasion process, which is numerically and analytically
 investigated in several limits. Generically advection enhances the
 downstream invasion speed but decreases the population size of the invading
 species, and can even inhibit the invasion process. Remarkably,
 however, the rate of population increase, which quantifies the
 invasion efficiency, is maximized by an optimal advection
 velocity. In  models with Allee effect, differently from the logistic case,
 above a critical unfavorable patch size the population localizes in a
 favorable patch, being unable to invade the habitat. However, we show
 that advection, when intense enough, may activate the invasion
 process.
\end{abstract}


\end{frontmatter}

\section{Introduction \label{sec:1}}

Invasions of alien species are widespread phenomena, in principle
affecting every ecosystem, usually with dramatic consequences on the
native community, constituting a major threat to biodiversity
\citep{Vitousek1997,Mooney2000,Pimentel2000}.  At the scale of interest for
management purposes, i.e. the geographic scale, invasive species move
across a heterogeneous landscape characterized by favorable and
unfavorable areas. The presence of abiotic heterogeneity, in fact,
characterizes most of natural habitats and plays a key role in
invasion processes, influencing their rate of spread and outcome
\citep{Shigesada1997,Hastings2005,Melbourne2007}.

Alongside with the empirical interest for the problem, several
modeling efforts have been dedicated to the understanding and
prediction of the spatial spread of invading organisms in
heterogeneous environments. Within the framework of reaction diffusion
models, building on the pioneering theoretical works of
\citet{Skellam1951} and \citet{Kierstead1953} on the ``critical patch
size'' problem, \citet{Shigesada1986} gave a seminal contribution
considering the invasion (propagation) of a population through a
periodic heterogeneous environment \citep[see
  also][]{Weinberger2002,Kinezaki2003}. The problem was extended
including advective transport to study persistence and propagation of
passively dispersing populations in oceans
\citep{Mann1991,Abraham1998} or rivers
\citep{Speirs2001,Pachepsky2005,Lutscher2006}. The importance of the
interplay between heterogeneity and advection has been recently
reviewed by \citet{Ryabov2008}. Moreover, the role of both advection
and landscape spatial structure is clearly relevant also to the
dispersal of plants \citep{Hastings2005}, whose seeds are transported
by winds.

In this article, we focus on the interplay between abiotic
heterogeneity and advection in invasions.  We describe the dynamics in
terms of an advection-reaction-diffusion model, which allows for
mathematical tractability and quantitative predictions, e.g., on the
spreading rates.  

We consider an infinite system where a population stably saturates the
carrying capacity on one side of the system and possibly invades the
remaining part of the environment, which is assumed to be
heterogeneous.  Our setting is quite general and widely applicable. In
particular, it is relevant to situations in which one has a
practically infinite biological reservoir of a species invading an
empty territory characterized by abiotic heterogeneity.  For instance,
the above setting may be relevant to situations in which invasions can
suddenly become possible for the removal of a climatic barrier due to
climate changes \citep{Mooney2000}.  Another relevant case is when a
species stably populating a lake invades an effluent characterized by
a certain degree of heterogeneity and stream velocity.  This is one of
the key early-stage processes related to the spatial control of
invasions in lakes' networks \citep{Havel2002}.  Other examples
concern the spreading of wind-pollinated plants in a heterogeneous
environment \citep{Davis2004} or spores carried by the wind
\citep{Kot1996}.

More specifically, the  habitat consists of a periodic
alternation of unfavorable and favorable patches, as in
\citet{Shigesada1986}. The population dies at a fixed rate in 
unfavorable regions, and grows in  favorable ones according to
either a logistic or an Allee effect dynamics. We are interested in
determining the conditions for invasions to be possible and in
understanding how invasion speed and efficiency depend on the
mechanisms at play.

With the logistic dynamics, in the absence of advection, this problem
was pioneered by \citet{Shigesada1986}, while \citet{Lutscher2006}
considered both advection and heterogeneity in reference to the
``drift paradox'' problem \citep{Speirs2001}.  Going beyond these
works, we find asymptotic expressions for both the invasion speed and
the rate of increase of the population size. The latter quantity
essentially estimates the rate at which the number of invading
individuals grows and, thus, provides a suitable measure of the
efficiency of the spreading process.  Indeed, especially in invasive
species control, it is important to quantify the potentiality of
growth of an alien population, and not only the speed at which it
colonizes the territory.  We anticipate that, remarkably, larger
invasion speeds do not necessarily imply more efficient invasions.

The logistic case (decreasing {\it per capita} growth rate) is then
contrasted with the case of positive density dependence corresponding
to a demographic Allee effect \citep{Allee1938,Dennis1989}, which
accounts for a reduced reproductive power at low densities. The
importance of Allee effects for the invasion and control of non-native
species was emphasized by \citet{Taylor2005} and \citet{Tobin2011}. It
is interesting to mention that even in homogeneous habitats the
presence of the Allee effect can decrease the invasion speed or even
halt the population spreading if the initially occupied area is too
small \citep{Lewis1993} (see also \citet{Vercken2011} for recent field
observations). We find that the interplay between heterogeneity and
advection becomes very subtle in the presence of the Allee effect.  In
fact it may happen that a persisting population, unable to invade new
territory, becomes able to spread in the presence of strong advection.
This effect should be taken as a cautionary note from the standpoint
of controlling invasive species, telling us that advective transport
should be considered.  For instance, after strong weather events, or
in regions characterized by prevailing winds, neglecting the effects
of advection could lead to the erroneous prediction of a population
unable to invade, whereas it actually propagates over the territory.

The material is organized as follows. In Sect.~\ref{sec:2} we present
the model, and in Sect.~\ref{sec:3} we qualitatively discuss its
phenomenology. Sections~\ref{sec:4} and \ref{sec:5} present and
discuss the main results on invasions with the logistic and the Allee
effect model, respectively. Finally, in Sect.~\ref{sec:6} we summarize the
results.

\section{Model \label{sec:2}}

The evolution of the population, $\theta(x,t)$, is governed by the
advection-reaction-diffusion equation
\begin{equation}
\partial_t \theta + v \partial_x \theta= D \partial^2_{x}\theta+f(\theta,x)\,.
\label{eq:ard}
\end{equation}
The diffusion coefficient $D$ and the advection velocity $v$ are
assumed to be constant.  Heterogeneity is introduced in the growth
term, $f(\theta,x)$, which depends on the position $x$.  The habitat
consists of a periodic alternation of unfavorable and favorable
patches of sizes $\ell_u$ and $\ell_f$, respectively. In the
elementary cell $[0:\mathcal{L}]$ (where $\mathcal{L}=\ell_u+\ell_f$
denotes the spatial period), we take
\begin{equation}
f(\theta,x)= \left\{
\begin{array}{ll}
g_u(\theta)  & \quad        \;0\leq x<\ell_u  \\
g_f(\theta)  & \quad \;\ell_u\leq x<\mathcal{L}
\end{array}
\right.\,.
\label{eq:react_hetero}
\end{equation}
In the unfavorable regions the population is assumed to die at a
constant rate $r_u$, so that $g_u(\theta)=-r_u\theta$ ($r_u>0$).  In the
favorable regions we consider two  dynamics.  The first is the
classical logistic model (with carrying capacity normalized to one)
\begin{equation}
g_f(\theta)=r_f\theta(1-\theta)\,,
\label{eq:logistic}
\end{equation}
$r_f$ being the intrinsic growth rate.  Equation~(\ref{eq:ard}) with
the logistic term but without advection was firstly studied by
\citet{Shigesada1986}.  Recently, \citet{Lutscher2006} included
advection focusing on the ``drift paradox'' problem.

Secondly, accounting for a positive correlation between population
density and {\it per capita} growth rate at small densities --- the
Allee effect~\citep{Allee1938,Dennis1989} ---, we consider the
threshold model
\begin{equation}
g_f(\theta)=r_f\max\{(\theta-\theta_c)(1-\theta),0\} \,,
\label{eq:allee}
\end{equation}
prescribing that the population grows only when $\theta>\theta_c$
(otherwise it stays constant). Notice that (\ref{eq:allee}) recovers
(\ref{eq:logistic}) for $\theta_c=0$. We remark that the model
(\ref{eq:allee}) represents an intermediate case between weak and
strong Allee effect~\citep[see also Sect.~\ref{sec:5} for further
  discussions]{Courchamp2008}. To the best of our knowledge,
 models with Allee effects have been mostly investigated in homogeneous
habitats \citep{Petrovskii2003}. In heterogeneous habitats we are
aware of only a few studies with integro-difference models
incorporating different dispersal kernels \citep[see, e.g., the recent
  work by][]{Dewhirst2009,Pachepsky2011}.  

We now specify the settings in which Eq.~(\ref{eq:ard}) is studied.
We consider model (\ref{eq:ard}) with boundary condition
$\theta(0,t)=1$, mimicking the case in which on the left of the origin
($x<0$) the population constantly saturates the carrying capacity, while the
population is initially absent in the $x>0$ region,
i.e. $\theta(x,0)=0$ for $x>0$.  With this choice for the boundary
conditions the invasion process must be considered from left to right
(i.e. from the biological reservoir at $x\leq 0$ to the positive real axis). In
this case depending on the sign of the advection velocity we can
consider (downstream) invasions with the flow (i.e when $v>0$) or
(upstream) invasions against the flow (i.e. when $v<0$).
\begin{figure*}[t!]
\centering
\includegraphics[width=1\textwidth]{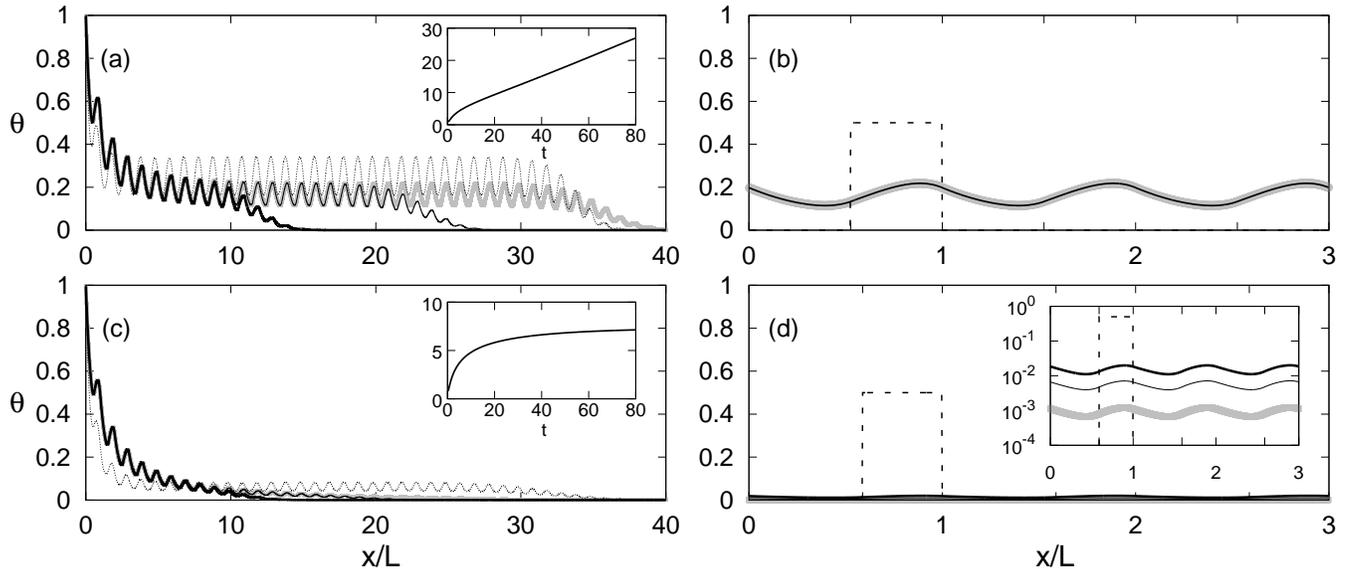}
\caption{Population evolution with logistic dynamics and boundary
  condition (BC) $\theta(0,t)=1$ (a,c) or with periodic BC,
  $\theta(0,t)=\theta(nL,t)$ with $n=3$(b,d).  Panel (a): Successful
  invasion for $l_u\!=\!2$, $l_f\!=\!1.8$, $\epsilon\!=\!1$ and
  $u=0.7$. The solid curves represent $\theta(x,t)$ at three
  successive times while the dotted curve corresponds to the case
  without advection ($u=0$), for comparison.  The inset shows
  $\int_0^\infty dx \theta(x,t)$ vs $t$.  Panel (b): Using the same
  parameters as in (a) but with periodic BC. The dashed rectangle
  corresponds to the initial condition, while the gray thick line is
  the population at stationarity. The black line superimposed on the
  gray one is obtained taking a cut of the interior of the traveling
  front from (a).  Panel (c): Unsuccessful invasion obtained in the
  same settings as in (a) but for $l_f=1.4$, notice that here for
  $u=0$ the invasion is still possible, meaning that advection is
  responsible for the halt of the invasion process.  The three solid
  curves refer to the same time instants of (a), while the dotted one
  is the $u=0$ case.  Inset as in (a).  Panel (d): parameters as in
  (c) but with periodic BC. Now the population goes extinct. The inset
  displays $\theta(x,t)$ at three successive times (from top to
  bottom) showing that $\theta\to 0$ exponentially.
\label{fig:figure1} }
\end{figure*}

It is useful to formulate the model in non-dimensional variables.  To
this aim we exploit known results about the logistic growth model
without advection, namely for the standard FKPP equation
\citep{Fisher1937,Kolmogorov1937}. The FKPP equation develops
traveling fronts characterized by the propagation speed $v_0=2\sqrt{Dr_f}$
and  width $\xi_0=\sqrt{D/r_f}$.  It is then natural to measure
lengths in units of $\xi_0$, time in units of the inverse growth rate
in the favorable patches $1/r_f$, and the advection velocity in units of
$v_0$. We thus define the non-dimensional variables $x' = x/\xi_0$,
$t' = t r_f$, $u=v/v_0$.  Dropping the primes, Eq.~(\ref{eq:ard}) made
non-dimensional reads
\begin{equation}
\partial_{t}\theta+2u\partial_{x}\theta=
\partial^2_{x}\theta+ f(\theta,x)\,.
\label{eq:ard-nd}
\end{equation}
The factor $2$ in the advection term results from our choice to fix
$u=1$ as the non-dimensional propagation speed in the
homogeneous FKPP system. We can now introduce $\epsilon=r_u/r_f$ which
is the death over growth rate ratio, and $l_{f,u}=\ell_{f,u}/\xi_0$
which are the non-dimensional sizes of the patches
($L=\mathcal{L}/\xi_0=l_u+l_f$). In this way, with reference to
Eq.~(\ref{eq:react_hetero}) we have $g_u(\theta)=-\epsilon \theta$
and
\begin{equation}
g_f(\theta) =\theta(1-\theta)\,,
\label{eq:react-nd}
\end{equation}
for the logistic model, while with an Allee effect it becomes
\begin{equation}
g_f(\theta)=\max\left\{(\theta-\theta_c)(1-\theta),0\right\} \,.
\label{eq:allee-nd}
\end{equation}
\begin{figure*}[t!]
\centering
\includegraphics[width=1\textwidth]{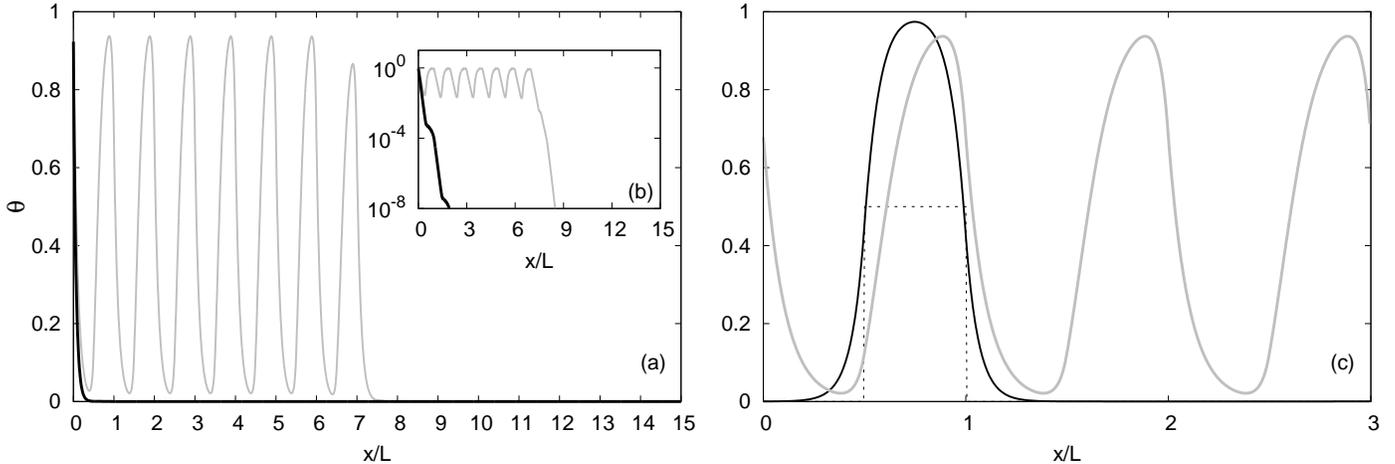}
\caption{Population evolution with the Allee effect model (\ref{eq:allee-nd}).
  The black and gray curves are obtained without advection ($u=0$) and
  with advection ($u=0.5$), respectively. The other parameters are
  $\epsilon=1$, $l_f=l_u=8$ and $\theta_c=0.001$. The different panels
  refer to: (a) BC $\theta(0,t)=1$, (b) the same of (a) but in
  logarithmic scale, (c) periodic BC with $n=3$ elementary cells.  The
  dashed line in (c) is the initial condition.
\label{fig:figure2} }
\end{figure*}

\section{Model phenomenology \label{sec:3}}
We now present the basic phenomenology of the model, discussing also
the main differences between logistic and Allee effect growth models. A
successful invasion implies the development, far from the boundary, of
a traveling front, characterized by a stationary and spatially
periodic bulk (Fig.~\ref{fig:figure1}a).  In such a case the total
population in the invaded habitat, $\int_0^\infty dx
\theta(x,t)$, asymptotically increases linearly with time (inset of
Fig.~\ref{fig:figure1}a). Faster growing populations mean more effective
invasions.  Conversely, Fig.~\ref{fig:figure1}c shows a typical case of
unsuccessful invasion: no traveling front develops and the total
population remains bounded in the limit of long times (compare the
insets in Figs.~\ref{fig:figure1}a and~\ref{fig:figure1}c).

For positive advection velocities, the problem of identifying the
conditions for successful invasions is directly related to determining
under which conditions Eq.~(\ref{eq:ard-nd}) with
(\ref{eq:react-nd}) in a finite system (of size $nL$ with $n$ integer)
with periodic boundary conditions ($\theta(0,t)=\theta(nL,t)$) admits
a non-vanishing stationary solution, starting from a generic non-zero
initial condition.  This is clearly shown by the perfect superposition
of the bulk of the traveling front with the stationary solution of the
periodic boundary condition problem (Fig.~\ref{fig:figure1}b).  The reason
for this link is that the bulk region of the traveling front (which is
stationary and spatially periodic) satisfies the same boundary value
problem of the finite system with periodic boundary conditions (BC).
Therefore, to determine whether downstream invasions are successful it
is enough to study whether persistence is possible in the finite
system with periodic BC. Conversely, an unsuccessful invasion implies
that in the finite system the population goes extinct, exponentially
in time (inset of Fig.\ref{fig:figure1}d).  

As we will see in the following, for small patch sizes, the
qualitative behavior of the different growth models is very similar:
invasions benefit from larger favorable regions and their speed
increases accordingly; moreover, the presence of advection enhances
the downstream invasion speed but decreases the population size of the invading
species, eventually halting the invasion (see Fig.\ref{fig:figure1}).
Conversely, for large enough unfavorable patches, dramatic differences
appear. In the Allee effect model, the population can persist in the absence
of advection but localized in a region of finite size, if it initially
occupied that area (see Fig.~\ref{fig:figure2}c), being unable to
invade new territories.  This is quite different from the logistic
model where invasions and persistence are always linked.  Even more
striking is the role of advection. Figures~\ref{fig:figure2}a,b show
that suitable values of the advection velocity can activate the
invasion of an otherwise localized population.

In the following we present the results for the logistic and Allee effect model
separately as the level of analytical understanding is quite
different. The possibility to use the linear analysis framework in the
logistic model, indeed, allows us to systematically derive the
conditions for invasions and asymptotic expressions for the invasion
speed and efficiency. This approach cannot be used for the Allee effect model, 
which is studied mainly numerically and with heuristic arguments.

\section{Results for the logistic growth model \label{sec:4}}

\subsection{Persistence  in a closed periodic system \label{sec:4.1}}
In this section we focus on the conditions for persistence in a closed
system with periodic BC, which correspond to those for successful
downstream invasion.  Moreover, when the population is able to
persist, we study how its size behaves as a function of the advection
velocity. It is worth noticing that the closed system setting is
interesting also in consideration of recent experiments where
bacterial populations are grown in heterogeneous conditions
\citep{Dahmen2000,Lin2004,Perry2005}.

\subsubsection{Critical patch size and critical advection \label{sec:4.1.1}}

Starting from a population different from zero in a single favorable
patch, with periodic BC, Eq.~(\ref{eq:ard-nd}) with
(\ref{eq:react-nd}) admits either
the trivial solution $\theta=0$, meaning that the habitat is unable to
sustain the population, or an asymptotically stationary non-vanishing
solution, when persistence is possible. To determine the conditions
for the latter, it is sufficient to identify when the solution
$\theta=0$ becomes linearly unstable, as briefly sketched in
\ref{app:1}.

For $l_u$ and $\epsilon$ fixed it is possible to show the existence of
both a critical size of the favorable patch, $l_f^*$, such that
extinction occurs if $l_f\leq l_f^*$, and a critical advection velocity
$u_c$ such that for a range of values of $l_f$ the system goes
extinct if $u>u_c$. The implicit relation between the critical values
reads
\begin{equation}
\begin{array}{l}
{\cosh(uL)} \!-\! \cos(\sqrt{1-u^2}l_f)\cosh(\sqrt{\epsilon+u^2}l_u) =
\\ \frac{\epsilon\!-\!1+2u^2}{2\sqrt{(1\!-\!u^2)(\epsilon\!+\!u^2)}}\sin(\sqrt{1\!-\!u^2}l_f)\sinh(\sqrt{\epsilon\!+\!u^2}l_u)
\end{array}.
\protect{\label{osdepv1}}
\end{equation}
The above equation was also found by \citet{Lutscher2006} using
boundary conditions different from ours.  Notice that for $|u|>1$ the
term $\sqrt{1-u^2}$ becomes imaginary, and in Eq.~(\ref{osdepv1}) the
identities $\sin(iz)=i\sinh(z)$ and $\tan(iz)=i\tanh(z)$ must be
employed. We also remark that Eq.~(\ref{osdepv1}) is left unchanged by
the substitution $u\to -u$ due to the symmetries of model
(\ref{eq:ard-nd}) with periodic BC. Therefore, we can limit the
analysis to $u\geq 0$.

\begin{figure}[t!]
\centering
\includegraphics[width=0.47\textwidth]{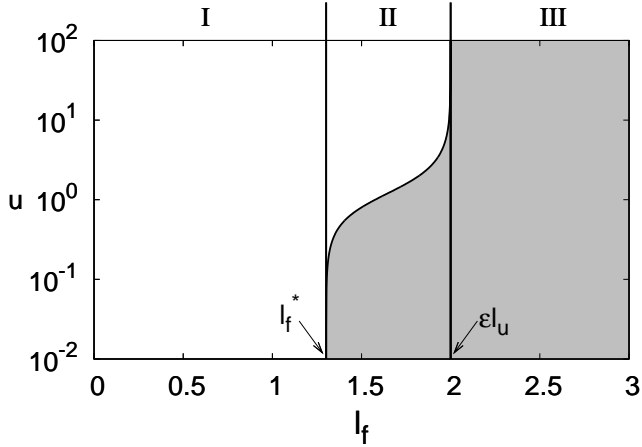}
\caption{Extinction (white) and survival (gray) regions in the plane
  ($l_f,u$).  The boundary curve between the two regions gives the
  dependence of the critical velocity $u_c$ on $l_f$.  The two
  vertical lines pinpoint the values $l_f^*$ and $\epsilon l_u$.  The
  Roman numbers on top label the three regions discussed in the
  text. Data refer to $l_u\!=\!2$ and
  $\epsilon\!=\!1$.  \label{fig:figure3}}
\end{figure}

Concerning the critical size of the favorable patch,
in the case of small unfavorable regions 
($l_u\ll 1$) the advection has not a great effect: at the leading order,
$l_f^* \approx \epsilon l_u$, i.e. the dependence on $u$ is
negligible.  However, for large unfavorable patch sizes, we have (see
also \citet{Speirs2001,Ryabov2008})
$$
\lim_{l_u \to \infty} l_f^* = \frac{2}{\sqrt{1-u^2}} \arctan \sqrt{\frac{\epsilon+u^2}{1-u^2}}\,,
$$ showing that advection worsens the survival conditions.
This effect can be deduced noticing that the advection term
changes the growth/death rate into
\begin{equation}
\epsilon(x)\to \epsilon(x)-u^2 \,, \label{eq:blasius}
\end{equation}
where $\epsilon(x)$ is the spatially dependent growth rate, taking the
values $1$ and $-\epsilon$ in the favorable and unfavorable patches,
respectively: essentially the effective growth rate is decreased by
$u^2$ while the effective death rate is increased by the same amount.
Equation~(\ref{eq:blasius}) can be derived from (\ref{eq:ard-nd}) via
the transformation $\theta(x,t) \!\to\!  \theta(x,t) e^{ux}$
\citep{Dahmen2000,Ryabov2008}. However, this transformation changes
the value of the density at the boundaries making the solution of the
periodic BC case more cumbersome.

When $l_f >l_f^*$, the population can be driven to extinction by
intense advection, exceeding a critical velocity $u_c$ that can be
computed from Eq.~(\ref{osdepv1}). Figure~\ref{fig:figure3} shows in
gray the region in parameter space $(l_f,u)$ where the population is
able to survive, for $\epsilon$ and $l_u$ fixed.  We can identify
three regions (as labeled on the top of Fig.~\ref{fig:figure3}): (I)
for $l_f \leq l_f^*$ the population goes extinct for any value of $u$;
(II) for $l_f^* < l_f \leq \epsilon l_u$ survival is possible below a
critical velocity $u_c$; for $l_f\to \epsilon l_u$ we have that $u_c
\to \infty$; (III) for $l_f > \epsilon l_u$ the average growth rate is
positive and the population survives for any value of $u$.

\subsubsection{Effects of advection velocity on the population size 
\label{sec:4.1.2}}
Now we study how the size of the population depends on the advection
velocity in order to characterize the transition from survival to
extinction and to derive some results to be used later
(Sect.~\ref{sec:4.2}).  In particular, we are interested in the
behavior of the average biomass defined as
\begin{equation}
B=\lim_{t\to\infty}\langle\theta(x,t)\rangle \equiv \lim_{t\to\infty}
\frac{1}{L}\int_0^{L}\theta(x,t){\mathrm d}x\,,
\label{eq:biomass}
\end{equation}
in the limit of large times, when the solution is stationary.  In the
above expression, thanks to the periodicity of $\theta(x,t)$, we
considered the biomass present in an elementary cell.  Given the habitat
properties, the biomass is a function of the advection velocity $u$,
$B=B(u)$.

\begin{figure}[bht]
\centering \includegraphics[width=0.45\textwidth]{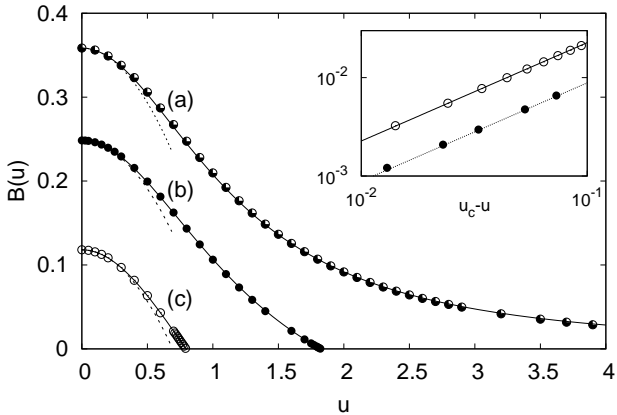}
\caption{Biomass $B$ vs $u$ with $l_u=2$ and $\epsilon=1$. Curve (a)
  corresponds to $l_f=\epsilon l_u=2$, at the border between region II
  and III; the other curves refer to values of $l_f$ in region II:
  $l_f=1.8$ (b) and $l_f=1.5$ (c).  Symbols show the results of
  simulations while the solid curves shows the fitting functions: for
  (a) Eq.~(\ref{eq:asymptvcinfty}) with $\beta=0.726$; for (b) and (c)
  Eq.~(\ref{eq:totalfit}) with $u_c=1.823,\, \beta=0.626$ and
  $u_c=0.794,\,\beta=0.494$, respectively.  The theoretical values for
  $u_c$ obtained from (\ref{osdepv1}) are $1\%$ close to the fitted
  values.  The dotted parabola shows the behavior
  (\ref{eq:bsmallu}). Inset: $B(u)$ vs $u_c-u$ for (b) and (c) close
  to the transition. The straight lines show the slopes $2B(0)/(u_c
  (1+\beta u_c^2))$ obtained from (\ref{eq:totalfit}) when $u\to
  u_c$. \label{fig:figure4}}
\end{figure}

With periodic boundary conditions, $B$ is an even function of $u$,
$B(u)=B(-u)$, so that, assuming that $B(u)$ is a smooth function, for
$u\ll 1$ we expect
\begin{equation}
B(u)=B(0) \left[ 1- \eta u^2+O(u^4) \right]\,,
\label{eq:bsmallu}
\end{equation}
with $\eta$ some positive constant, as confirmed by numerical
simulations (dotted curves in Fig.~\ref{fig:figure4}). The biomass
decreases with $u$ because the net effect of advection is to
increase/decrease the death/growth rate, as from
Eq.~(\ref{eq:blasius}).  Equation~(\ref{eq:bsmallu}) agrees with
results of \citet{Dahmen2000} for the linearized dynamics. Even though
they also claim that with the complete equation non analytic behaviors
(i.e. $B(0)-B(u)\propto |u|$) may appear due to the
nonlinearity. However, our simulations always confirmed (\ref{eq:bsmallu}).

The behavior (\ref{eq:bsmallu}) holds both in region II and III of
Fig.~\ref{fig:figure3}. In region III, where survival is possible for
any value of the advection velocity, we have that in the limit $u\to
\infty$ the average biomass attains a finite limiting value given by
(see \ref{app:2}) $ B(u) = \Delta/l_f + O(\Delta/u^2)\,, $ where
$\Delta=l_f-\epsilon l_u$, is nothing but the average growth rate
times $L$. Curve (a) in Fig.~\ref{fig:figure4} shows $B(u)$ at the
transition between region II and III, i.e. for $\Delta=0$. In this
case $u_c=\infty$, so that $B(u\to\infty)=0$ and the function
\begin{equation}
B(u)=\frac{B(0)}{1+\beta u^2}
\label{eq:asymptvcinfty}
\end{equation}
provides a very good fit of $B(u)$ for any value of $u$.  Notice that
Eq.~(\ref{eq:asymptvcinfty}) implies (\ref{eq:bsmallu}) with
$\eta=\beta$.

In region II, the critical velocity $u_c$ is finite and, by
definition, $B(u) \to 0$ when $u \to u_c$. As typical in phase
transitions, we should expect $B(u)\sim (u_c-u)^\nu$ for $u_c-u\ll 1$,
with $\nu$ some exponent characterizing the extinction
transition. Assuming a smooth behavior it is reasonable to expect
$\nu=1$, as confirmed by the inset of Fig.~\ref{fig:figure4} and
supported by analytical approaches by \citet{Dahmen2000} valid in the
limit $l_u\to \infty$. Finally, assuming the simplest functional form
consistent with the symmetries and regularity properties of $B(u)$ we
end up with the expression
\begin{equation}
B(u)=B(0)\frac{1-\left(u/u_c\right)^2}
              {1+\beta u^2}\,,
\label{eq:totalfit}
\end{equation}
which is consistent with (\ref{eq:bsmallu}) for $u\ll 1$ giving
$\eta\!=\!\beta+u_c^{-2}$, and with (\ref{eq:asymptvcinfty}) for
$u_c\to \infty$. Curves (b) and (c) in Fig.~\ref{fig:figure4} show
$B(u)$ for two values of $l_f$ within region II. For both values one
can observe the very good agreement between numerical data and
Eq.~(\ref{eq:totalfit}). The above results show to what extent
advection decreases the average population size, and provide a
characterization of the advection-induced population extinction.

\subsection{Effects of heterogeneity and advection on the invasion speed and efficiency\label{sec:4.2}}

As discussed in Sect.~\ref{sec:3}, Eq.~(\ref{osdepv1}) also provides
the condition for downstream invasions (i.e. when $u>0$) from a
reservoir (on the left) to a heterogeneous habitat (on the right, as
in Fig.~\ref{fig:figure1}a).  As for upstream invasions (i.e. when
$u<0$), it is necessary to understand when Eq.~(\ref{eq:ard-nd})
admits solutions which develop a periodic traveling front advancing
with a positive speed $u_p$, for long times and far from the boundary.
In the following, we show how the speed $u_p$ can be derived, and
discuss the general conditions for invasions.

Assuming that, far from the boundaries, a traveling front develops,
following \citet{Shigesada1986} we can write
$\theta(x,t)=\Theta(z)g(x)$, where $z=x-2u_pt$ accounts for
propagation with velocity $u_p$ (the factor $2$ deriving from our
choice of the non-dimensional variables, see Sect.~\ref{sec:2}).  The
function $\Theta(z)$ describes the traveling front modulated by a
periodic function $g(x)=g(x+L)$ due to the habitat periodicity.  For
the computation of the invasion speed we can use the linearized dynamics
assuming that the traveling component has an exponential leading edge
$\Theta(z) \propto e^{-sz}$. As detailed in \ref{app:3}, we end up
with an implicit relation between the invasion speed $u_p$ and the
shape parameter $s$ of the traveling front (see also
\citet{Lutscher2006})
\begin{eqnarray}
\mathcal{D}(u_p,s;u) & = & \cosh(q_0L)  
 -  \cosh(q_uL_u)\cosh(q_fl_f) \nonumber \\
& - &\frac{q_u^2+q_f^2}{2q_uq_f}\sinh(q_ul_u)\sinh(q_fl_f) = 0 \label{eq:disprelfunc}
\end{eqnarray}
where $q_0=u+s$, $q_u=\sqrt{u^2+\epsilon+2su_p}$ and
$q_f=\sqrt{u^2-1+2su_p}$.  From (\ref{eq:disprelfunc}) one derives
$u_p(s;u)$ and the invasion speed can be obtained by computing
$\min_{s}\{u_p(s;u)\}$.  As far as we know, there is no analytical
expression for the minimum, and numerical computations must be
employed. In the sequel, with some abuse of notation we will denote
with $s(u)$ the value of $s$ for which the minimum is realized and
with $u_p(u)$ the minimal speed.

\begin{figure}[htb]
\centering
\includegraphics[width=0.45\textwidth]{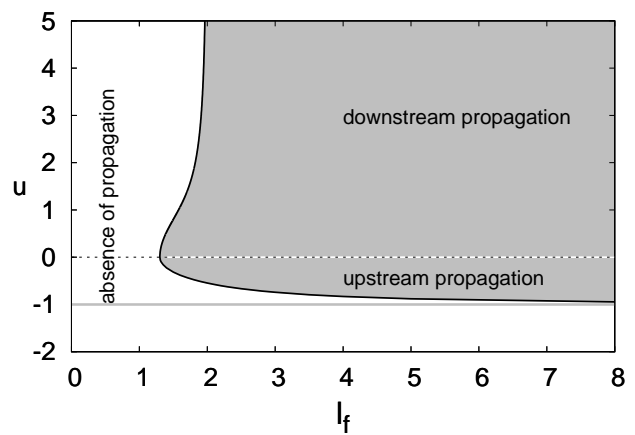}
\caption{Regions where invasions are successful (gray) and
  unsuccessful (white) in the plane ($l_f,u$). For $u>0$ we have
  downstream invasions. For $u<0$, upstream invasions are possible
  only for $-1<u<0$, see text for explanation. Data refer to the case
  $l_u=2$ and $\epsilon=1$.\label{fig:figure5}}
\end{figure}

The gray area in Fig.~\ref{fig:figure5} displays the region in the plane
($l_f,u$) where invasions are possible. Such a region was numerically
determined by solving Eq.~(\ref{eq:disprelfunc}) and finding the
values of $l_f$ and $u$ for which the propagation speed
$u_p(u)=\min_{s}\{u_p(s;u)\}$ exists and is positive.  For $u>0$ the
curve, separating white and gray regions, coincides with that derived
from Eq.~(\ref{osdepv1}) (shown in Fig.~\ref{fig:figure3}).  For $u<0$, it
approaches the asymptote $u=-1$ for large $l_f$. Indeed for very large
favorable patches the system should recover the homogeneous habitat
result $u_p=1+u$, so that invasions are impossible for $u<-1$
\citep{Lutscher2006}.

We now focus on the invasion speed $u_p$ for downstream invasions
($u>0$). The upstream case was considered in details by
\citet{Lutscher2006}.  Simulations (not shown) 
suggest that for $u\ll 1$ the invasion speed behaves linearly in $u$,
i.e.
\begin{equation}
u_p(u)\approx u_p(0)+\alpha u\,,
\label{eq:smallu}
\end{equation}
both in region II and III.
The above result holds also for small negative $u$.  To determine
$\alpha$ the first step is to expand $\mathcal{D}(u_p,s;u)$ in powers
of $u$. At the first order the expansion yields
\begin{eqnarray}
\mathcal{D}(u_p,s;u) =  \mathcal{D}(u_p,s;0) + f(s)u+O(u^2) =0 \label{eq:disprelfirstorder}\,,
\end{eqnarray}
with $f(s)=L\sinh(Ls)$.  For small $u$ the minimum of
(\ref{eq:disprelfirstorder}) is realized at $u_p(u)=u_p(0)+\delta
u_p$ and $s(u)=s(0)+\delta s$, with $\delta u_p, \delta s \sim O(u)$. 
Expanding now Eq.~(\ref{eq:disprelfirstorder}) in $\delta u_p$
and $\delta s$, one  gets
\begin{equation}
\delta u_p=-\left[\frac{f(s(0))}{\left.\partial_{u_p} \mathcal{D}(u_p,s;0)\right|_{u_p(0),s(0)}}\right]\,u=\alpha u\,.
\label{eq:deltavf}
\end{equation}
From the numerical values of $u_p(0)$ and $s(0)$, we computed the
value of $\alpha$ obtaining a perfect agreement with simulations.

\begin{figure}[th]
\centering
\includegraphics[width=0.45\textwidth]{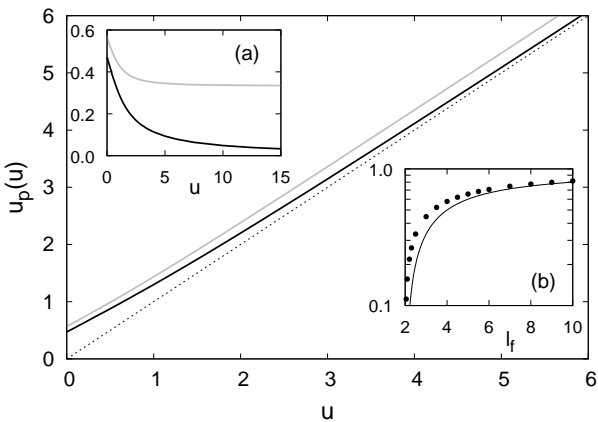}
\caption{Invasion speed $u_p$ vs  $u$ for $l_f=2$
  (black curve) and $l_f=2.5$ (gray curve), together with the linear
  behavior $u_p=u$ (dotted line). Inset (a): $u_p-u$ vs $u$ for
  $l_f=2$ (black curve) and $l_f=2.5$ (gray curve). Inset (b): $u_p-u$
  vs $l_f$ (symbols) compared with the asymptotic prediction
  (\ref{eq:up_ulargeu}), approximated in the simulation using $u=100$.
  Data have been obtained with $l_u=2$ and $\epsilon=1$.}
\label{fig:figure6}
\end{figure}

We now consider the behavior of the invasion speed for large advection
velocities. When $l_f$ is chosen in region II
 the invasion is halted for $u>u_c$. In region III,
the population can invade the habitat for any $u$ and the invasion
speed approaches another linear behavior for $u\gg 1$.  In
Fig.~\ref{fig:figure6}, we contrast two cases: when $l_f=\epsilon
l_u$ and $l_f>\epsilon l_u$, inside region III.

As discussed in Sect.~\ref{sec:4.1.2}, when $l_f=\epsilon l_u$, the
average biomass vanishes ($B(u)\to 0$) for $u\to u_c=\infty$.  It is
thus reasonable to expect that the contribution to the invasion speed
comes only from advection. Therefore, asymptotically we expect $u_p
\to u$ as shown in Fig.~\ref{fig:figure6}, though the convergence of
$u_p-u$ to zero can be rather slow
(Fig.~\ref{fig:figure6}a). Conversely, inside region III, $u_p-u$
reaches a finite value for $u\to\infty$ (Fig.~\ref{fig:figure6}a).  As
heuristically derived in \ref{app:4}, for $u\gg 1$ one expects
\begin{equation}
u_p(u)-u={\Delta}/{l_f}\,,
\label{eq:up_ulargeu}
\end{equation}
where $\Delta=l_f-\epsilon l_u$.  Strictly speaking, the above result
holds in the limit of $u\to\infty$ and $l_f\to\infty$ but, as shown in
Fig.~\ref{fig:figure6}(b), it is in fairly good agreement with the numerical
results also for finite values of $l_f$.

To summarize, in region III the invasion speed $u_p$ is well approximated 
by two different linear behaviors
\begin{equation}
\protect{\label{eq:summary}}
\left \{
\begin{array}{ll}
u_p(u) \simeq u_p(0)+\alpha u & \quad \mbox{for small } u\vspace{0.2truecm}\\
u_p(u) \simeq {\displaystyle \Delta}/{\displaystyle l_f} + u & \quad \mbox{for large } u\end{array} \right .
\end{equation}
where $\alpha$ is given by
Eq.~(\ref{eq:deltavf}).

\subsubsection{Invasion speed in rapidly and slowly varying environments \label{sec:4.2.1}}

We now study the invasion speed in two limiting cases for which some
analytical results can be obtained, namely when the habitat is finely
fragmented ($l_{u,f}\ll 1$) or subdivided in large patches
($l_{u,f}\gg 1$ with $\gamma=l_f/l_u$ fixed).

When $l_{f,u} \ll 1$, expanding Eq.~(\ref{eq:disprelfunc}) at the
lowest order, one obtains the explicit expression
\begin{equation}
u_p(s;u) \approx \frac{s}{2} + \left(\frac{\Delta}{L}\right) \frac{1}{2s} + u\,,
\label{eq:reldisp_1ord}
\end{equation}
Retaining higher order terms, it is possible to show that the
dependence of $u_p$ on $u$ is linear up to the fourth order in
$l_{f,u}$, where a term proportional to $u^2$ appears.
Minimizing Eq.~(\ref{eq:reldisp_1ord}) yields
\begin{equation}
u_p(u) =  \sqrt{{\Delta}/{L}} + u\,,
\label{eq:vf_v1ord}
\end{equation}
as also obtained with homogenization techniques~\citep{Lutscher2006}.
Equation~(\ref{eq:vf_v1ord}) shows that for finely fragmented habitats
the intrinsic propagation speed is as in the homogeneous habitat once
the growth rate is substituted with the average growth rate
$\Delta/L$, as found by \citet{Shigesada1986} for $u=0$. Moreover,
Eq.~(\ref{eq:vf_v1ord}) implies that for $l_{u,f} \to 0$, $\alpha$ in
Eq.~(\ref{eq:smallu}) is equal to unity. We finally observe that
Eq.~(\ref{eq:vf_v1ord}) holds only for $\Delta\geq0$, that is in
region III of Fig.~\ref{fig:figure3}. Indeed for $l_{u,f} \to 0$
region II shrinks to zero.
 
In the  limit of very large patch sizes, $l_{f,u} \rightarrow
\infty$, from Eq.~(\ref{eq:disprelfunc}) we obtain
\begin{equation}
q_0 L  = q_ul_u + q_fl_f \,,
\label{eq:reldisp_infty_log}
\end{equation}
where $q_0\!=\!u\!+\!s$, $q_u\!=\!\sqrt{u^2\!+\!\epsilon\!+\!2su_p}$
and $q_f\!=\!\sqrt{u^2\!-\!1\!+\!2su_p}$ (see \ref{app:3}).  It is
interesting to compute the limit maintaining the ratio $\gamma=l_f/l_u$
constant so that Eq.~(\ref{eq:reldisp_infty_log}) becomes $q_0
(1+\gamma) = q_u + \gamma q_f$, which has the explicit form
\begin{eqnarray}
(1\!+\!\!\gamma) (u \!+\! s) \!=\! \!\sqrt{u^2
    \!+\!\!\epsilon \!+\! 2u_ps} \!+\!\! \gamma
   \sqrt{u^2\!\!-\!1\!+\!2u_ps}. \label{eq:reldsip_infty_vf}
\end{eqnarray}
The above formula does not depend on $l_{u}$ and $l_f$
separately but only on their ratio $\gamma$, meaning that for $l_{u,f}
\to \infty$ the propagation speed $u_p$ approaches a limit value that
depends only on $\gamma$ and $\epsilon$.
The limits $\gamma \gg 1$ and $\gamma \ll 1$ are quite trivial and
consistent with intuition.  In both cases, neglecting sub-leading
terms, squaring both sides of~(\ref{eq:reldsip_infty_vf}) one finds
$u_p(s)$. In the former limit, $l_f \gg l_u$ (negligible unfavorable
patches), the homogeneous result $u_p(u)=1+u$ is retrieved. In the
latter, $l_u \gg l_f$ (negligible favorable patches), the condition
for the minimum of $u_p(s;u)$ is realized for imaginary
values of $s$, meaning that invasions are not possible.
\begin{figure}[t!]
\centering
\includegraphics[width=0.45\textwidth]{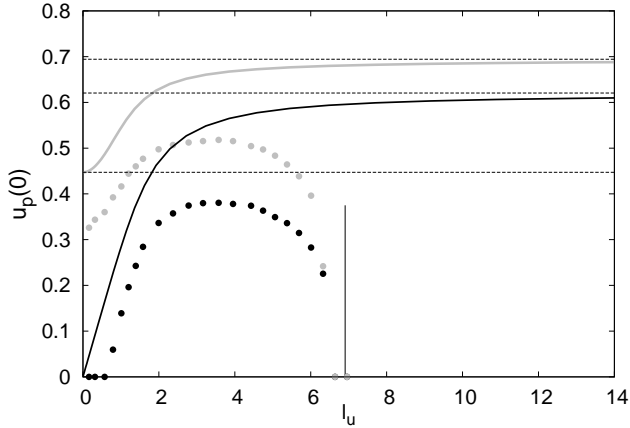}
\caption{Invasion speed $u_p(0)$ vs $l_u$ with $\gamma=l_f/l_u$ fixed
  and $\epsilon=1$. Continuous lines correspond to the numerical
  solution of Eq.~(\ref{eq:disprelfunc}) for $\gamma=1$ (black) and
  $\gamma=1.5$ (gray). Dashed lines correspond to the limit values for
  rapidly and slowly varying environments, respectively. The former is
  zero for $\gamma=1$. Symbols show $u_p$ measured in numerical
  simulations with the Allee effect model (\ref{eq:allee-nd}) with
  $\theta_c=0.001$ (see Sect.~\ref{sec:5}). Black and gray symbols
  correspond to $\gamma=1$ and $1.5$, respectively. The vertical bar
  marks the critical unfavorable patch size (\ref{eq:ligni}) above
  which invasions are impossible.\label{fig:figure7}}
\end{figure}

In the special case $\gamma\!=\!1$ and $u\!=\!0$,
Eq.~(\ref{eq:reldsip_infty_vf}) becomes $2s = \sqrt{2u_p
  s+\epsilon} + \sqrt{2u_ps - 1}$, from which it is easy to derive the
invasion speed:
\begin{equation}
u_p(0)= 2\frac{1+\epsilon^2+(1-\epsilon)\sqrt{1+\epsilon+\epsilon^2}}{\left(1-\epsilon+2\sqrt{1+\epsilon+\epsilon^2}\,\right)^{3/2}}.
\label{eq:vfstar}
\end{equation}
Notice that, once the proper correspondence between notations is made,
the result (\ref{eq:vfstar}) coincides with that obtained by~\citet{Hamel2010}
using a different technique.  Further specializing to the case of equal
growth and death rates ($\epsilon=1$) Eq.~(\ref{eq:vfstar}) reduces to
$u_p(0) = 2^{1/2}3^{-3/4}$.

In the presence of advection with $u\ll 1$, the computation for
$\gamma=1$ and $\epsilon=1$ can be easily extended obtaining
$u_p(u)=2^{1/2}3^{-3/4}+(2/3)u$ that is Eq.~(\ref{eq:smallu}) with
$\alpha=2/3$. For $u\gg 1$, we were unable to obtain analytical
results, but we expect the phenomenology discussed in Fig.~\ref{fig:figure6}
to apply.

In Fig.~\ref{fig:figure7} we show the invasion speed $u_p$, obtained by
numerically solving (\ref{eq:disprelfunc}) for $u=0$, at varying
$l_{u,f}$ with $\gamma=l_f/l_u=1$ and $1.5$.  We also show the
asymptotic values for $l_{u,f} \to 0$ and $l_{u,f}\to \infty$. In
rapidly varying environments $u_p(0)=\sqrt{(\gamma -
  \epsilon)/(1+\gamma)}$ while in slowly varying ones $u_p(0)$ goes to
the finite value (\ref{eq:reldsip_infty_vf}) which for $\gamma=1$ is
given by (\ref{eq:vfstar}).

\subsubsection{Efficiency of  the invasion process as a function of the advection velocity}
The invasion speed $u_p$ measures the velocity of population
advancement. We now focus on the rate of increase of the population
size (inset of Fig.~\ref{fig:figure1}a), which provides a measure of
the efficiency of the invasion process.  The suitable quantity to look
at is the rate of increase of the total biomass
$$
B_r(u) = \frac{\mathrm d}{\mathrm d t} \int_0^\infty\theta(x,t){\mathrm d}x\,,
$$
which is the slope of the curve shown in the inset of Fig.~\ref{fig:figure1}a.
\begin{figure}[th!]
\centering
\includegraphics[width=0.45\textwidth]{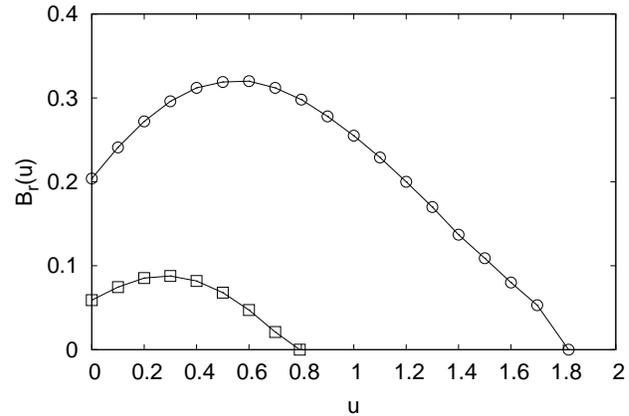}
\caption{Rate of increase of the biomass $B_r(u)$ vs   $u$ for
  $\epsilon=1$, $l_u=2$ and $l_f=1.5$ (boxes) and $l_f=1.8$ (circles)
  (see also Fig.~\ref{fig:figure4}). \label{fig:figure8}}
\end{figure}

Figure~\ref{fig:figure8} shows that there is an optimal advection velocity
which maximizes the rate $B_r(u)$. So that, in spite of the fact that
larger advection velocities imply larger invasion speeds, the
efficiency of the invasion process is maximized at a specific value
$u^\ast$ of the advection velocity.  This means that even though for
$u>u^\ast$ the invasion speed increases, the number of invading
individuals decreases, which implies a less effective invasion.

We now provide a heuristic argument to explain the origin of an
optimal advection velocity.  At stationarity, the population,
advancing at constant speed $u_p(u)$, increases its size at a rate
\begin{equation}
B_r(u) = \,B(u) u_p(u)
\label{eq:ratebiomassvv}
\end{equation}
where $B(u)$ is the average biomass (\ref{eq:biomass}).  For $u\ll 1$,
$B(u)$ is well described by Eq.~(\ref{eq:bsmallu}), i.e. $B(u)$
decreases quadratically with $u$. On the other hand, as from
Eq.~(\ref{eq:smallu}), the invasion speed increases linearly with $u$.
Using the above considerations and Eq.~(\ref{eq:ratebiomassvv}), we
obtain that the increase in invasion speed will dominate at very small
$u$ while the quadratic decrease of $B(u)$ will dominate at larger
$u$, producing the bell shaped behavior observed in
Fig.~\ref{fig:figure8}.

The above argument is based on a low order Taylor expansion which, in
principle, could cease to be valid for values of $u$ at which the
maximum of $B_r(u)$ is attained. We numerically found that the
behavior reported in Fig.~\ref{fig:figure8} is general and that,
typically, the maximum of $B_r(u)$ is realized for values of $u$ for
which the Taylor expansion is still a valid approximation.

\subsection{Discussions\label{sec:4.3}}

\citet{Shigesada1986} have shown that even when the average growth
rate is negative (i.e. $\Delta=l_f-\epsilon l_u<0$) there exists a
critical size of the favorable patches above which a population can
invade new territories. We have shown that the main effect of
advection is to increase the critical size $l_f^*$ for the invasion to
be possible. In particular, there always exists a critical advection
velocity $u_c$ above which no invasion is possible, unless the
average growth rate is positive.  

As for the invasion speed, we recover the results obtained for $u=0$
by \citet{Shigesada1986}, and for $u\neq 0$ by \citet{Lutscher2006},
who focused on upstream propagation (i.e. in our setting
$u<0$). However, here, we mainly focused on the downstream invasion
speed, i.e.  when the advection velocity favors the invasion
($u>0$). Our analysis shows that, provided $u<u_c$, advection always
increases the invasion speed. In particular, both for small and large
advection velocities $u$ the invasion speed is linear in $u$ but with
different prefactors (see Eq.~(\ref{eq:summary})).  In some
interesting environmental limits, moreover, we analytically computed
the invasion speed. Our results agree with those found with different
techniques by \cite{Hamel2010}, and are interesting in view of the
ensuing discussion on Allee effects.

Our results on the dependence of the population size (the biomass) on
advection extend similar ones derived by \citet{Dahmen2000} in the
limit of large unfavorable patches. Although the invasion speed is
enhanced by advection, the biomass decreases at increasing the
advection velocity. As a consequence, a faster invasion speed does not
necessarily imply a more efficient invasion process. Indeed, the
suitable quantity to judge about the effectiveness of the invasion
process is the rate of increase of the biomass. We found the
remarkable new result of an optimal advection velocity maximizing
such rate and, hence, the efficiency of the invasion process.  This
maximum originates from the opposite role of advection on the invasion
speed $u_p$ and on the biomass $B(u)$: the former increases linearly
with $u$ while the latter decreases quadratically with $u$.  The
balance between these two behaviors leads to a maximum in the invasion
efficiency.

\section{Results for the Allee effect growth model\label{sec:5}}

Some features of the logistic model seem quite unreasonable from an
ecological point of view. For instance, as shown in
Fig.~\ref{fig:figure7}, the invasion speed approaches an asymptotic
value when the sizes of the patches are enlarged holding fixed their
ratio, regardless of the size of hostile regions (i.e. even for
$l_u\to \infty$). Intuition would suggest that extremely large
unfavorable regions should slowdown, and eventually suppress,
invasions, as observed, e.g., in soil organisms \citep{Bailey2000}.  A
slowdown of the invasion process should be expected, indeed, anytime
there is a positive correlation between population density and {\it
  per capita} growth rate, i.e. in the presence of Allee  effects
\citep{Allee1938,Dennis1989}.  In general, one speaks of Allee effect
when for small densities the growth rate is negative --- strong Allee
effect --- or positive, but smaller than for larger density --- weak
Allee effect \citep{Wang2001}.  Relevant works and their relations
with our problem are discussed in the next section (in particular, see
Table \ref{tab:allee_models}).

We consider the Allee effect threshold model
(\ref{eq:allee-nd}), that we recall here
\begin{equation}
g_f(\theta)=\max\{(\theta-\theta_c)(1-\theta),0\} \,.
\label{eq:allee-nd2}
\end{equation}
This model corresponds to a situation in between the strong and weak
Allee effect  and is convenient
because it reduces to the logistic model (\ref{eq:react-nd}) for
$\theta_c\!=\!0$, easing the comparison.  At the end of this section
we will briefly discuss different Allee effect models.

Studying the problem without advection allows
us to identify the main consequence of the Allee effect, namely the
existence of a critical unfavorable patch size $l^*_u$ above which
invasion is impossible for any size of the favorable region and the
population can persist localized as shown in Fig.~\ref{fig:figure2}.  At
stationarity, in the bulk of the traveling front we have that, if
$\theta_M$ is the value at the beginning of an unfavorable patch, then
at the end of the unfavorable region the density will reach the value
$\theta_m=\theta_M\exp\left(-\sqrt{\epsilon}l_u\right)$.  Given the
growth term (\ref{eq:allee-nd2}), for the population to propagate we
must require that $\theta_m\geq \theta_c$ (see also
\citet{Dewhirst2009} for a similar argument applied to an
integro-difference model), which implies the inequality
\begin{equation}
l_u\le l^*_u=\frac{1}{\sqrt{\epsilon}} \log\left(\frac{\theta_M}{\theta_c}\right)
\label{eq:ligni}
\end{equation}
for the unfavorable patch size.  In general, $\theta_M$ cannot be
estimated analytically; however, setting $\theta_M\!=\!1$ gives a
reasonable upper bound.

The existence of $l^*_u$ is evident from Fig.~\ref{fig:figure7} where
symbols denote the results of numerical simulations obtained with the
growth model (\ref{eq:allee-nd2}), holding constant $\gamma=l_f/l_u$
and increasing the patch sizes. For small sizes the qualitative
behavior of the model (\ref{eq:allee-nd2}) is similar to that of the
logistic model: $u_p$ increases with $L$.  A dramatic difference
appears at large sizes: for the logistic model the invasion speed
reaches an asymptotic value, while for the Allee effect one it decreases and,
eventually, the invasion process is halted when $l_u \approx l_u^*$,
regardless the size of the favorable patch.

\begin{figure}[t!]
\centering
\includegraphics[width=0.45\textwidth]{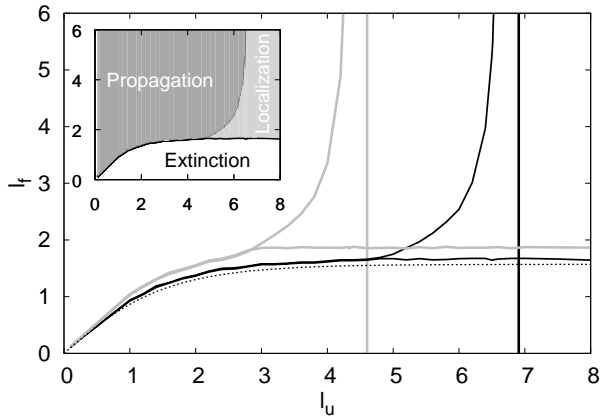}
\caption{Regions of the plane $(l_u,l_f)$ characterized by
  localization, propagation and extinction of the population with the
  Allee effect model (\ref{eq:allee-nd2}) with $u=0$, $\epsilon=1$,
  $\theta_c=0.001$ (black) and $\theta_c=0.01$ (gray).  For each value
  of $\theta_c$, two curves separate the three different possibilities
  as shown in the inset with shaded areas for the case
  $\theta_c=0.001$.  The vertical lines in the main figure show
  $l_u^*$ obtained from (\ref{eq:ligni}) with $\theta_M=1$. The dotted
  curve displays the dependence of $l_f^*$ for the logistic model
  (\ref{eq:react-nd}).
  \label{fig:figure9}}
\end{figure}

In Figure~\ref{fig:figure9}, we show the behavior of the system
without advection in the plane $(l_u,l_f)$, for two values of
$\theta_c$.  As already discussed, for the logistic growth model there
exists a critical favorable patch size $l_f^*$ above which the
population can survive and invade new territories.  The critical value
$l_f^*$ remains finite for $l_u \to \infty$ as found by
\citet{Ludwig1979,Shigesada1986} and also in this paper.  With the
Allee effect growth model (\ref{eq:allee-nd2}) the phenomenology is different
and more interesting. When the size of the unfavorable patches is
small (i.e. $l_u\ll 1$) the system essentially behaves as the
logistic model: the unfavorable patch is so small that typically the
decrease in density will not cause the population to fall below
$\theta_c$. From a quantitative point of view, $l_f^*$ is slightly
larger than the logistic value (the effect being more pronounced for
larger $\theta_c$), though this cannot be fully appreciated from
Fig.~\ref{fig:figure9} due to the scale.  The main qualitative change
with respect to the logistic case manifests when the size of the
unfavorable patches approaches the value $l_u^*$ given in
Eq.~(\ref{eq:ligni}).  For unfavorable patches at least this large,
propagation becomes impossible for any size of the favorable patches,
even though the population does not necessarily go extinct. Indeed, as
highlighted in the inset of Fig.~\ref{fig:figure9}, a new region in
the plane $(l_u,l_f)$ appears, where the population can persist
locally but cannot propagate (see Fig.~\ref{fig:figure2}): it
localizes in a single favorable patch (if initially it was in that
patch). For this localization regime to exist it is necessary that
$l_f$ is wide enough to sustain the population, as theoretically
derived in the homogeneous case by \citet{Lewis1993} and found in
field data by \citet{Vercken2011}.

Therefore, without advection but with the Allee effect growth term
(\ref{eq:allee-nd2}), we have that the population goes extinct if
$l_f<l_f^*$, propagates if $l_f>l_f^*$ but $l_u< l_u^*$, and localizes
in a single favorable patch when this is large enough to sustain the
population ($l_f>l_f^*$) but the unfavorable patch is too large to
allow the propagation, i.e. $l_u \geq l_u^*$.
\begin{figure}[t!]
\centering
\includegraphics[width=0.45\textwidth]{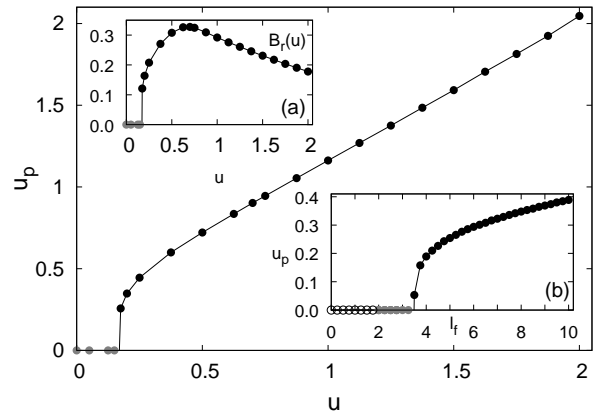}
\caption{Invasion speed $u_p$ vs $u$ for the Allee effect model
  (\ref{eq:allee-nd2}) with $\theta_c=0.001$, $\epsilon=1$ and
  $l_u\!=\!l_f\!=\!8$, i.e. within the localization region of
  Fig.~\ref{fig:figure9}.  Gray symbols correspond to values of $u$
  for which the population remains localized, while black ones to
  values for which it propagates.  Inset (a): rate of increase of the
  biomass $B_r$ vs $u$ obtained using the same parameters of the main
  figure.  Inset (b): invasion speed vs $l_f$ for $u=0.2$ and $l_u=
  8$. At increasing $l_f$ the population passes from extinction (empty
  symbols) to localization (gray symbols) and finally to propagation
  (black symbols). The last transition is possible only thanks to
  advection.
\label{fig:figure10}}
\end{figure}

We now discuss the effects of advection.  For small
unfavorable patch sizes, $l_u\ll 1$, since the model with Allee effect
behaves quite similarly to the logistic model, also the effect of advection
on the dynamics is very similar between the two models.
A critical advection velocity $u_c$ exists, above
which the population goes extinct. However, for larger unfavorable
patches, while in the absence of advection propagation is inhibited
and the population remains localized (light gray region in the inset
of Fig.~\ref{fig:figure9}), sustained advection induces the
remarkable qualitative changes observed in Fig.~\ref{fig:figure2}. In
Fig.~\ref{fig:figure10} we show a numerical measurement of the
invasion speed $u_p$ as a function of $u$ for a habitat with
unfavorable and favorable patch sizes, $l_u$ and $l_f$ respectively,
chosen in the localization region of the system without advection.  As
it can be seen, while the population remains localized at small
advection velocities, it suddenly becomes able to propagate invading
the whole environment when the velocity of the medium becomes large
enough.  An intuitive explanation for such a behavior is that, thanks
to advection, the population can now travel through the unfavorable
patch more rapidly, finally reaching the next favorable patch with a
density above the threshold $\theta_c$, i.e., advection enhances the
value $l_u^*$ above which no propagation is possible.  In fact, at
stationarity, in the unfavorable region it is easy to see that the
density behaves as $\theta(x)=\theta_M\exp[(u-
  \sqrt{\epsilon+u^2})x]$. Therefore, the same argument which lead to
Eq.~(\ref{eq:ligni}) now yields
$$
l_u^*(u)= \frac{1}{\sqrt{\epsilon+u^2}-u}\log\left(\frac{\theta_M}{\theta_c}\right)\,.
$$ The above formula predicts that $l_u^*(u)$ grows with $u$, so that
even if in a medium at rest the population is localized,
i.e. $l_u>l_u^*(0)$, in advective media there will be a value of the
advection velocity such that $l_u\leq l_u^*(u)$, allowing the
population to propagate, as shown in Fig.~\ref{fig:figure10}.  Once invasion
is permitted by advection, the propagation speed and the rate of
increase of biomass behave similarly to the same quantities in the
logistic model (compare Fig.~\ref{fig:figure10} main figure with
Fig.~\ref{fig:figure6} main figure, and Fig.~\ref{fig:figure10}a with
Fig.~\ref{fig:figure8}).
\begin{figure*}[t!]
\centering
\includegraphics[width=0.9\textwidth]{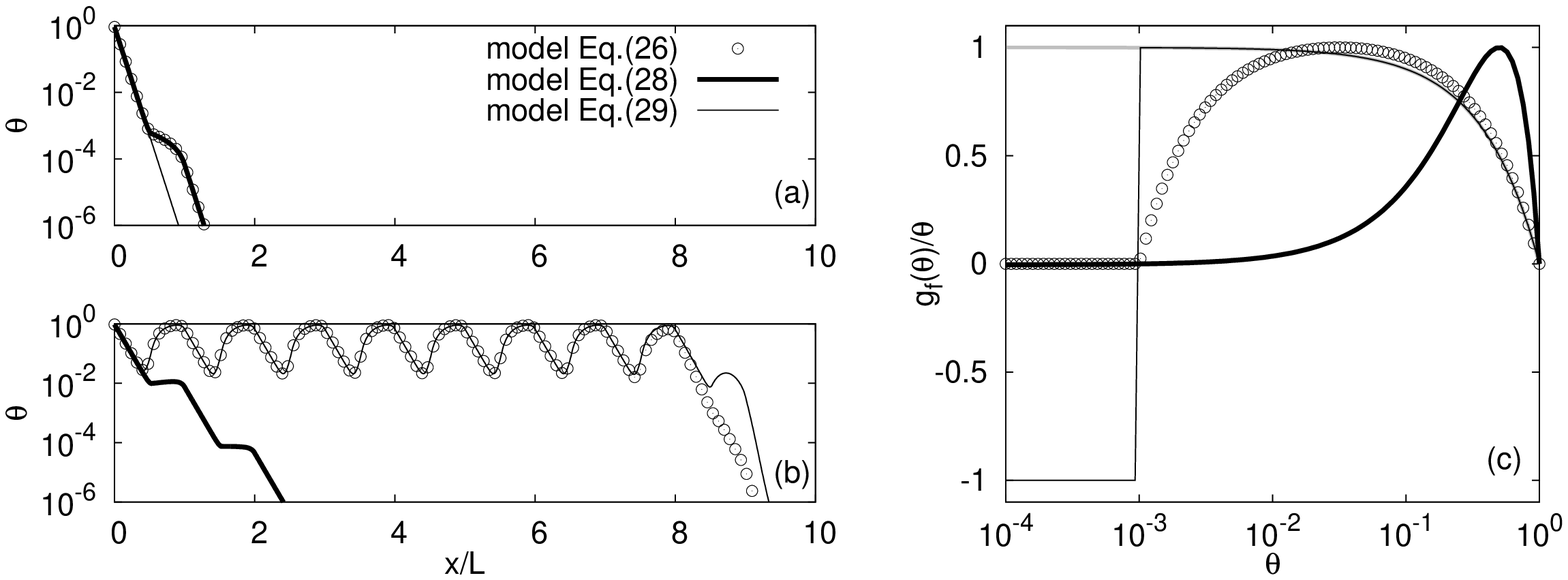}
\caption{Advection-induced invasions for different Allee effect models with
  $\theta_c\!=\!0.001$, $l_u\!=\!8$, $l_f\!=\!7$ and $\epsilon\!=\!1$.
  (a) With $u\!=\!0$ the population evolving with the Allee effect models as
  labeled is unable to invade. (b) With $u\!=\!0.5$ invasion is
  possible but for the cubic model (\ref{eq:cubic}). As for model
  (\ref{eq:wang}) we used $b\!=-1$. (c) Comparison of the
  per capita  growth rates $g_f(\theta)/\theta$ in the three
  considered models together with the logistic case
  $g_f(\theta)/\theta\!=\!1\!-\!\theta$ (gray line).  Notice that for
  comparison purposes all models have been normalized in a such a way
  that $\max_\theta\{g_f(\theta)/\theta\}=1$.
\label{fig:figure11}}
\end{figure*}

The results shown in the inset (b) of Fig.~\ref{fig:figure10} further
illustrate the importance of advection in such model. There, we show
the measured propagation velocity $u_p$ as a function of $l_f$, for
constant advection $u$ and keeping $l_u>l_u^*(0)$ fixed, so that in
the absence of advection the population would be localized even if
$l_f>l_f^*$.  As one can see two transitions are observed: from
extinction to localized, mainly due to the increase of $l_f$; from
localized to propagating, which is possible only thanks to advection.
It is interesting here to note that something similar happens if we
consider a spatially dependent diffusion coefficient, as done
by~\citet{Shigesada1986} for the logistic growth model. In particular,
without advection ($u=0$) and denoting with $\delta=D_u/D_f$ the ratio
between diffusivities in the unfavorable and favorable patches,
Eq.~(\ref{eq:ligni}) modifies in $l_u^* \approx \sqrt{\delta/\epsilon}
\ln(\theta_M/\theta_c)$.  This result tells us that a larger
diffusivity in the unfavorable regions ($\delta>1$) allows the
population to propagate also with larger unfavorable patch
sizes. Therefore, if we fix the size of the unfavorable patch, either
increasing the diffusivity or the advection, the residence time in the
unfavorable patch decreases leading to a smaller depletion in
population density.

Through this section we limited the numerical analysis to the Allee
effect model (\ref{eq:allee-nd2}) which is intermediate between the case of
weak and strong Allee effect. It is thus worth to test the robustness
of the above findings by considering different Allee effect models.
In Figure~\ref{fig:figure11} we show the evolution of invasions
obtained with three different models of the Allee effect, either in
the absence (Fig.~\ref{fig:figure11}a) or in the presence
(Fig.~\ref{fig:figure11}b) of advection when $l_u>l_u^*$. In
particular, we compare model (\ref{eq:allee-nd2}) with two models of
Allee effect, namely the standard cubic term
used to model strong Allee effects
\citep{Lewis1993,Petrovskii2003,Almeida2006}
\begin{equation}
g_f(\theta)=\theta(1-\theta)(\theta-\theta_c)\,,
\label{eq:cubic}
\end{equation}
and  a slight modification of the model introduced by
\citet{Wang2001}
\begin{equation}
g_f(\theta)=\left\{
\begin{array}{ll}
b\theta  & \quad        \;0\leq \theta \leq\theta_c  \\
\theta(1-\theta)  & \quad \;\theta_c< \theta\leq 1
\end{array}
\right. \,,
\label{eq:wang}
\end{equation}
The latter can model either strong ($b<0$) or weak (if $0<b<1$ and
$1-\theta_c>b$) Allee effect. The marginal case $b=0$, for small
values of $\theta_c$, is practically indistinguishable from the Allee
effect model (\ref{eq:allee-nd2}). Figure~\ref{fig:figure11}a shows that with
$u=0$ invasions are unsuccessful in all the considered models.  For
model (\ref{eq:wang}) we only show the case $b=-1$. However, the
result (not shown) holds also for positive (not too large) values of
$b$, meaning that also with weak Allee effects the population can be
localized.  Advection (Fig.~\ref{fig:figure11}b) is able to activate
the invasion process in all cases but for the cubic model
(\ref{eq:cubic}).  The origin of such a difference is readily
understandable from Fig.~\ref{fig:figure11}c where we show the
density-dependent per capita growth rate $g_f(\theta)/\theta$ in the above models
compared with the logistic case (the gray line). As one can see, the
growth rate for the cubic model, although positive, remains very small
for a large interval above the threshold value $\theta_c$. As a
consequence, even if the population reaches the favorable region with
$\theta>\theta_c$, it cannot propagate. In the other cases the growth
rate is large enough to allow the population to grow and propagate.

Concluding, the phenomenology of the advection-induced invasion is
rather general in the presence of (either weak or strong) Allee
effect, but critically depends on the growth rate realized close to
the threshold value $\theta_c$. In perspective, it would be
interesting to investigate what habitat/advection conditions
ensure successful invasions for different kinds of Allee effects.
\begin{table*}[t!]
\centering
\begin{tabular}{llll}
\hline
\hline
               & \textbf{habitat} & \textbf{dispersal}     & \textbf{references} \\
\hline
{\small invasion}       & {\small homogeneous}   & {\small diffusion}   & {\small Lewis and} \\
               &               & {\small (short range)}   & {\small van den Driessche (1993),} \\
               &               &                           & {\small Hastings (1996),} \\ 
               &               &                           & {\small Wang and Kot (2001)} \\
               &               &                           & \\
{\small invasion,}      & {\small homogeneous}   & {\small diffusion+advection}     & {\small Lewis and Kareiva (1993),} \\
{\small persistence,}   &               & {\small (short+long range)}      & {\small Petrovskii and Li (2003),} \\
{\small localization}   &               &                           & {\small Almeida et al. (2006)} \\          
               &               &                           & \\
{\small persistence}    & {\small heterogeneous} & {\small diffusion}                 & {\small Shi and Shivaji (2006)}\\ 
               &               & {\small (short range)}             & \\
               &               &                           & \\
{\small invasion,}      & {\small heterogeneous} & {\small diffusion+advection}     & {\small This work} \\
{\small localization,}  &               & {\small (short + long range)}       & \\
{\small persistence}    &               &                           & \\
               &               &                           & \\
\hline
{\small invasion,}       & {\small heterogeneous} & {\small dispersal kernel}           & {\small Dewhirst and} \\
{\small localization,}   &               & {\small (long range)}        & {\small Lutscher (2009),} \\ 
{\small persistence}    &               & {\small (special case: diffusion)}      & {\small Pachepsky and} \\
               &               &                           & {\small Levine (2011)} \\ 
\hline
\hline
\end{tabular}
\caption{Models with Allee effects related to the present work. The
  models in the upper part of the table are
  advection-reaction-diffusion models; those in the lower part of the
  table are integro-difference models. Further details about models of
  Allee effects in invasion dynamics can be found in tables $3.5$ and
  $3.6$ in \citet{Courchamp2008}. Our work aims at investigating
  conditions for persistence and invasion in heterogeneous
  environments with advection.}
\label{tab:allee_models}
\end{table*}

\subsection{Discussions\label{se:5.1}}

To put in perspective our results on the Allee effect model it is useful
first to briefly recall some known results from the literature.  The
study of Allee effects has received considerable interest in the past,
with growing modeling efforts in recent years, when its importance for
invasions and conservation issues have started to be well recognized
\citep{Taylor2005,Courchamp2008,Tobin2011}. For convenience of the
reader, we summarize relevant works related to our problem in Table
\ref{tab:allee_models}.

In the case of homogeneous environments and without advection, most
studies have shown that the success of an invasion depends not only on
the initial density, but also on the size of the initially occupied
area \citep{Lewis1993,Lewis_VDD1993,Kot1996,Wang2001}.  Asymptotic
rates of spread are typically reduced \citep[see, e.g.][]{Lewis1993},
mostly because of the interplay between dispersal mechanisms and the
reduced reproductive power at low densities. In discrete-space models
\citep[e.g., in][]{Keitt2001} the interesting phenomenon of population
localization (also called range pinning) occurs. The presence of
advection can lead to nontrivial results when combined with
density-dependent migration \citep{Petrovskii2003,Almeida2006}.

In the case of heterogeneous habitats, with few exceptions (see, e.g.,
\cite{Shi2006} who extended the critical patch size problem to the
case of weak Allee effects), most works focused on integro-difference
models \citep{Dewhirst2009}, sometimes accounting also for the
effect of discreteness of the population \citep{Pachepsky2011}. These
studies agree on the fact that the presence of Allee effects combined
with habitat fragmentation generally penalizes the success of
invasions or, at least, slows them down.  In particular, crucial for
the success of the invasion process is the size of bad patches
\citep{Dewhirst2009} with respect to the dispersal range.  Moreover,
the results of these models depend also on the choice of the
dispersal kernel, which is typically poorly known
\citep{Hastings2005}.

Our results show that logistic and Allee effect models qualitatively display
the same features both in the presence and in the absence of advection
when the unfavorable patches are not too large, including the
nontrivial existence of an optimal advection velocity maximizing the
invasion efficiency.  However, in the Allee effect case, unlike the logistic
model, if the unfavorable patches become too large invasions can be
halted.  When the Allee effect is included, in fact, if both $l_u$ and
$l_f$ are large, the population can persist locally, but is unable to
invade other patches, since it cannot cross unfavorable regions
without being too severely damped. This means that while in the
logistic case the thresholds for persistence and for invasions
coincide, these are in general different when Allee effects are
present.  The same observation was made in the context of
integro-difference models \citep{Dewhirst2009}, and it is
substantiated also by field observations (\citet{Bailey2000}, see also
\citet{Hastings2005} and references therein).

Advection, however, can alter this picture: if its intensity exceeds a
threshold value, the invasion process can be activated, provided that
a large enough growth rate is realized at the population density when
entering the favorable patches.  This peculiar effect of advection,
which as far as we know was not previously put into light, has obvious
ecological implications for invasive species management strategies.
For instance, the idea to induce Allee effects to control the invasion
of alien species, which can be implemented in several ways
\citep{Tobin2011}, should be pursued with extreme care in the presence
of advection. 

In summary, our results on persistence and invasion in the presence of
Allee effects extend previous works done in the framework of
reaction-diffusion models (with homogeneous habitat) and positively
correlate also with the results of integro-difference models,
emphasizing the subtle role of advection which was missed in that kind
of models.

\section{Conclusions \label{sec:6}}

In this paper we focused on the role of advection on invasions in
heterogeneous environments characterized by favorable and unfavorable
patches.

On the one hand, we have shown that in the presence of habitat
heterogeneity sufficiently intense advection can halt the invasion
process. Moreover, we argued that the efficiency of the invasive
process is properly quantified in terms of the rate of increase of the
invading population. In particular, we found that the latter is
maximal at intermediate values of the advection velocity.

On the other hand we have shown that in the presence of Allee effects,
advection may be beneficial to the invasion process turning a
persistent but non invading population into an invading one.

An important aspect in evaluating biological invasions, which has not
been considered in our work, is related to the discrete nature of a
population, which is made of individuals
\citep{Durrett1994,Okubo2001}.  We expect that considering also
discrete effects may add further nontrivial effects due to demographic
stochasticity, in particular close to the transition between
successful and unsuccessful invasions. Works in these direction have
started to appear \citep{Snyder2003,Pachepsky2011}. It would be
interesting to study the effect of advection also in the presence of
demographic stochasticity.

\section{Acknowledgments}
We thank A. Vulpiani, who contributed to the initial stage of this
work, for fruitful discussions and interactions.
MC acknowledges support from MIUR PRIN-2009PYYZM5
``Fluttuazioni: dai sistemi macroscopici alle nanoscale''.

\appendix
\section{Conditions for persistence in a periodic system\label{app:1}}

Due to the habitat periodicity, and with the periodic BC, we can limit
the analysis to the unit cell $[0:L)$. Denoting with $\vartheta$ a
  perturbation around the solution $\theta=0$, we have that
  $\vartheta$ is ruled  by the linearized version of
  Eq.~(\ref{eq:ard-nd})
\begin{equation}
\partial_t \vartheta +2u\partial_x \vartheta=\partial^2_x \vartheta +\epsilon(x)\vartheta\,,
\label{eq:vartheta}
\end{equation} 
where $\epsilon(x)=-\epsilon$ in the unfavorable patches and $\epsilon(x)=1$
in the favorable ones.
In the linear analysis framework, at leading order, one expects that
$\vartheta= e^{\Lambda t}\psi(x)$ so that population extinction is a
stable solution if $\Lambda<0$ and an unstable one if
$\Lambda>0$. When unstable, the asymptotic solution will be a stationary and
spatially periodic solution $\psi(x)$ as in Fig.~\ref{fig:figure1}a,
otherwise if $\Lambda<0$ it decays exponentially as
$\vartheta(x,t)=\psi(x)e^{-|\Lambda| t}$ (see inset of
Fig.~\ref{fig:figure1}c). Plugging $\vartheta= e^{\Lambda t}\psi(x)$ into
(\ref{eq:vartheta}) we obtain the characteristic equation for the
stationary state
\begin{equation}
  \partial^2_{x}\psi - 2u \partial_x \psi +(\epsilon(x)-\Lambda)  \psi  = 0\,,
\label{eq:app}
\end{equation}
whose general solution is given by
\begin{equation}
\begin{array}{ll}
\psi_{u}(x) = A^u_1 e^{x\lambda^{(u)}_{1}}+A^u_2e^{x\lambda^{(u)}_{2}}
&  \;   0\leq x < l_u\\ 
\psi_{f}(x) = A^f_1 e^{x\lambda^{(f)}_{1}}+A^f_2 e^{x\lambda^{(f)}_{2}} 
&  \; l_u\leq x < L\,,
\end{array}
 \label{eq:t}
\end{equation}
where 
\begin{equation}
\begin{array}{l}
\lambda^{(u)}_{1,2}= u\pm\phantom{i}
\sqrt{\epsilon+\Lambda+u^2}=u\pm b_u(\Lambda) \\
\lambda^{(f)}_{1,2}=u\pm
i\sqrt{1-\Lambda-u^2}=u\pm i b_f(\Lambda) 
\end{array}
\label{eq:eigen}
\end{equation}
are the eigenvalues associated to Eq.~(\ref{eq:app}). For a solution
to exist it is sufficient to impose the continuity of densities
$\psi_{u,f}(x)$ and fluxes $J_{u,f}(x)=
[2u-\partial_{x}]\psi_{u,f}(x)$ at the boundaries between favorable
and unfavorable regions. Notice that the conditions on fluxes
are required when using space-dependent diffusion coefficient, as, for
example, in \citet{Shigesada1986} and \citet{Lutscher2006}.  For
constant diffusion coefficient, as here, it is enough to impose the
continuity of derivatives of $\psi$.

Using Eq.~(\ref{eq:t}) and imposing the aforementioned continuity conditions
we obtain a linear system for the four constants $A^{u,f}_{1,2}$. Requiring
that this linear system has a nontrivial solution we obtain 
\begin{eqnarray}
&&\cosh(uL) -
\cos(b_f(\Lambda)l_f)\cosh(b_u(\Lambda)l_u)
\nonumber \\
&&=\frac{b_u^2(\Lambda)-b_f^2(\Lambda)}{2b_u(\Lambda)
b_f(\Lambda)} \sin(b_f(\Lambda)l_f)\sinh(b_u(\Lambda)l_u) \,.\label{eq:relation2}
\end{eqnarray}
The largest value 
of $\Lambda$ solving the above equation determines
the stability properties of the solution $\theta=0$.  In particular,
fixing the values of $l_u$, $\epsilon$ and $u$, it is possible to show
that $\Lambda\geq 0$ whenever $l_f\geq l_f^*$, where $l_f^*$ solves
Eq.~(\ref{eq:relation2}) with $\Lambda=0$
\citep{Shigesada1986,Nagylaki1975}, i.e.  Eq.~(\ref{osdepv1}).
For $l_f<l_f^*$ the population always goes extinct. An equivalent
critical value exists for the advection velocity, see main text.  Let
us notice that with $u=0$, the above condition reduces to that found
by \citet{Shigesada1986}, i.e.  $\sqrt{\epsilon}\tanh\left(
  \sqrt{\epsilon} {l_u}/{2} \right )= \tan\left ( {l_f}/{2} \right
)\,$.  For $l_u\ll 1$, the above equation tells us that $l_f^* \approx
\epsilon l_u$. Notice that the average growth rate is given by
$\left(l_f-\epsilon l_u\right)/L$ and that for $l_f > \epsilon l_u$
survival of the population is guaranteed by the fact that the average
growth rate is positive. As the size of the unfavorable patches grows
the critical (favorable) patch size approaches the limit value $l_f^*
= 2 \arctan\sqrt{\epsilon}$ corresponding to the result of
\citet{Ludwig1979}. The case $\epsilon \to \infty$, i.e. infinite
mortality, corresponds to the KISS critical patch size $l_f^*=\pi$
\citep{Skellam1951,Kierstead1953}.

\section{Derivation of the asymptotic expression for biomass.\label{app:2}}
At stationarity, for $u\gg 1$ we can
disregard the diffusive term in Eq.~(\ref{eq:ard-nd}) obtaining:
\begin{equation}
\theta^{\prime}=\left\{
\begin{array}{ll}
-{\epsilon\theta}/{(2u)} & \quad 0\leq x< l_u \\
{\theta(1-\theta)}/{(2u)} &\quad l_u\leq x<L 
\end{array}
\right.
\label{eq:largeu}
\end{equation}
where the prime represents the derivative with respect to $x$. Then,
denoting with $\theta_M=\theta(0)$ and $\theta_m=\theta(l_u)$ the
values of $\theta$ at the beginning of the unfavorable and favorable
regions, respectively (which correspond to the maximum and minimum
realized values of $\theta$), Eq.~(\ref{eq:largeu}) is solved by
\begin{equation}
\label{eq:largeuS}
\theta(x)=\left\{
\begin{array}{ll}
\theta_M \exp\left(-\frac{\epsilon x}{2u}\right) &   0\leq x< l_u \\
\noalign{\vskip 2mm}
\frac{\theta_m \exp\left(\frac{x-l_u}{2u}\right)}{1-\theta_m\left[1-\exp\left(\frac{x-l_u}{2u}\right)\right]} & l_u\leq x<L 
\end{array}
\right.\,.
\end{equation}
Now imposing the periodicity $\theta_M=\theta(L)$ and noticing that
$\theta_m=\theta_M \exp(-\epsilon l_u/(2u))$ we find that
\begin{equation}
\theta_M= \frac{\exp\left(\frac{\Delta}{2u}\right)-1}{\exp\left(\frac{\Delta}{2u}\right)-\exp\left(\frac{\Delta-l_f}{2u}\right)}
\label{eq:tetaM}
\end{equation}
where $\Delta=l_f-\epsilon l_u$.  The above expression provides a
meaningful solution only for $\Delta>0$, i.e.  in region III where
extinction never takes place.  Moreover, for $u\to\infty$
Eq.~(\ref{eq:tetaM}) can be expanded to show that it reaches a finite
limit $\theta_M=\Delta/l_f(1+O(1/u))$.  Integrating (\ref{eq:largeuS})
in $[0:L]$ one obtains an explicit expression for $B(u)$, which
expanded for large $u$ gives $ B(u) = \Delta/l_f + O(\Delta/u^2)\,.  $

\section{Derivation of the dispersion relations\label{app:3}}
For long times, as shown in Fig.~\ref{fig:figure1}, the bulk of the
traveling front is a periodic function, $
\theta(x,t)=\theta(x+L,t+T)\,, $ where $L$ coincides with the habitat
spatial period and $T$ is the temporal period; the invasion
(propagation) speed is then given by $L/T$. With the chosen BC the
propagation proceeds in the positive $x$ direction. To derive the
propagation speed we can write $\theta(x,t)=\Theta(z)g(x)$ with
$z=x-2u_pt=x+L-2u_p(t+T)$ and $g(x)=g(x+L)$ being a periodic function
which modulates the traveling front \citep{Shigesada1986}.  We
can now take $\Theta(z)\propto e^{-sz}$ meaning that, apart from the
periodic modulation $g(x)$, the leading edge is exponentially
decaying. Plugging the above expressions in the linearized
Eq.~(\ref{eq:ard-nd}) yields the equation for $g$
\begin{equation}
g^{\prime\prime}\!\!-\!2(u+s) g^\prime \!\!+\![\epsilon(x)+2(u\!-\!u_p)s+s^2]g \!=\!0\,.\label{appeq:eqforg}
\end{equation}
 If $\epsilon(x)=1$, $g$ is constant and the homogeneous case is
 recovered. The linear equation (\ref{appeq:eqforg}) has the general
 solution $g_u(x) =c_{1}^u e^{\mu^{(u)}_{1} x}+c_{2}^u
 e^{\mu^{(u)}_{2} x}$ and $g_f(x) =c_{1}^f e^{\mu^{(f)}_{1} x}+c_{2}^f
 e^{\mu^{(f)}_{2} x}$ in the unfavorable and favorable patches,
 respectively and
\begin{eqnarray}
\mu^{(u)}_{1,2} &=&
u+s\pm\sqrt{u^2+\epsilon+2su_p} =q_0\pm q_u \nonumber\\
\mu^{(f)}_{1,2} &=&
u+s\pm\sqrt{u^2-1+2su_p}=q_0\pm q_f \,.\nonumber
\end{eqnarray}
Imposing the continuity conditions as in \ref{app:1}, we find four
equations for the coefficients $c_{1,2}^{f,u}$, and requiring the
existence of a nontrivial solution, we find that the dispersion
relations (\ref{eq:disprelfunc}) must be satisfied. Notice that
(\ref{eq:disprelfunc}) was also found by \citet{Lutscher2006} using
different BC. Moreover for $u=0$ Eq.~(\ref{eq:disprelfunc}) reduces to
the equation found by \citet{Shigesada1986}.

\section{Invasion speed in the limit of large advection\label{app:4}}
The values of $u_p(u)-u$ and $\theta_M$ (the maximum value of the
population density) can be written in general as $u_p(u)-u=f(u,
\epsilon, l_f, l_u)u_0$ and $\theta_M=g(u, \epsilon, l_f,
l_u)\theta_0$, where $u_0=1$ and $\theta_0=1$ are the values in a
homogeneous environment.  Taking the ratio between the above equations
we obtain:
$$u_p(u)-u=\frac{f(u, \epsilon, l_f, l_u)} {g(u, \epsilon, l_f,
  l_u)}\theta_M\,.$$ Now, if we conjecture that the decrease of the
difference $u_p(u)-u$ with $u$ is mainly controlled by the dependence
of $\theta_M$ on $u$, we can make the strong assumption that $f(u,
\epsilon, l_f, l_u) / g(u, \epsilon, l_f, l_u) \approx 1$, which is
true at least for $l_f \to \infty$ as both functions tend to
$1$. Using Eq.~(\ref{eq:tetaM}) of \ref{app:2} in the limit
$u\to\infty$, which implies $\theta_M=\Delta/l_f$, we finally obtain $
u_p(u)-u={\Delta}/{l_f}\,.  $ This equation is true in the limit of
$u\to\infty$ and $l_f\to\infty$ but also for finite values of $l_f$ it
is in fairly good agreement with the numerical results (see Fig.~\ref{fig:figure6}b).

\end{document}